\documentclass[aps,prd,twocolumn, showpacs,nofootinbib]{revtex4}
\usepackage{amsmath}
\usepackage{graphicx}
\usepackage{dcolumn}
\usepackage{bm}
\usepackage{amssymb}
\usepackage{latexsym}

\newcommand{\no}{\nonumber}

\newcommand{\pa}{\partial}
\newcommand{\del}{\delta}
\newcommand{\eps}{\epsilon}

{\bf }

\newcommand{\F}{{\cal F}}

\newcommand{\half}{\frac{1}{2}}

\newcommand{\Del}{\Delta}

\begin{document}

\title{Quantum speed limit for  relativistic spin-0 and spin-1 bosons on commutative and noncommutative planes}

\author{Kang Wang ${}^a$}

\author{Yu-Fei Zhang ${}^a$}

\author{Qing Wang ${}^b$}

\author{Zheng-Wen Long ${}^c$}





\author{Jian Jing ${}^{a}$}
\email{jingjian@mail.buct.edu.cn}

\affiliation{${}^a$ Department of Physics and Electronic, School of
Science, Beijing University of Chemical Technology, Beijing 100029,
P.R. China,}

\affiliation{${}^{b}$ College  of Physics and Technology, XinJiang University, Urumqi 830046, P.R.China}

\affiliation{$^{c}$ College of Physics, GuiZhou University, GuiYang, 550025, P.R. China}

\begin{abstract}
Quantum speed limits 
of  relativistic charged spin-0 and spin-1 bosons in the background of a homogeneous magnetic field are studied on both commutative and noncommutative planes. 
We show that, on the commutative plane, 
the average speeds of wave packets along the radial direction during the interval 
in which a quantum state evolving from an initial state to the orthogonal final one can not exceed the speed of light, 
regardless of the intensities of the magnetic field. 
However, due to the  noncommutativity, the  average speeds of the wave packets on noncommutative plane will exceed the speed of light in vacuum provided the intensity of the magnetic field is strong enough. It  is 
a clear signature of violating  Lorentz invariance in quantum mechanics region. 



\end{abstract}


{} 


\maketitle

\section{Introduction}

Duffin-Kemmer-Petiau (DKP) equation is a first-order relativistic wave equation \cite{Petiau, Duffin, Kemmer}. Different from Dirac equation which describes spin-$\half$ fermions, DKP equation describes spin-0 and spin-1 bosons. DKP equation takes the form
\begin{equation}
(i \beta^\mu \pa_\mu - m_0) \psi=0, \label{dkp}
\end{equation}
where $m_0$ is the rest mass and the matrices $\beta^\mu$ satisfy the algebraic ralation
\begin{equation}
\beta^\mu \beta^\nu \beta^\alpha + \beta^\alpha \beta^\nu \beta^\mu = g^{\mu \nu} \beta ^\alpha + g^{\alpha \nu} \beta^\mu. \label{cbeta}
\end{equation}
Here, $g^{\mu \nu} =g _{\mu \nu} = {\rm diag} (+,-,-,-)$ is the metric tensor. The algebra (\ref{cbeta}) has three different representations: a (one-dimensional) trivial representation, a five-dimensional representation describing  spin-0 bosons and a ten-dimensional representation describing  spin-1 bosons. As a Dirac-type exactly solvable relativistic quantum mechanical model,  DKP equation is studied from various aspects in the past years.
The magnetic coupling in DKP equation is also considered  \cite{nowakowski}. When magnetic field is taken into consideration, one should introduce the magnetic potentials  by the minimal coupling (we set $\hbar=c=1$),
\begin{equation}
p_i \to p_i + q A_i, \label{mc}
\end{equation}
where $q$ and $A_i$ are  the charge and the magnetic potentials respectively. 
Since DKP equation is analogous with Dirac equation,  the author of Ref. \cite{nowakowski} compares it with Dirac equation in detail in that paper.


On the other hand, the minimum time of a quantum state evolving from an initial state to the orthogonal final one in Hilbert space is of great importance in the field of quantum computation, quantum control and quantum metrology. In fact, it has attracted  attention  for a long time \cite{mt}. At the present, there are two different descriptions of the minimum time for a quantum system evolving from an initial state to the orthogonal final state. One is given by the expression $T_{\rm min} = \frac{\pi \hbar}{2 \Del E}$, in which $\Del E$ is the energy variance, defined by  $\Del E = \sqrt {\langle \psi |H ^2| \psi \rangle -\langle \psi |H |\psi \rangle ^2}$, with $H$ being the Hamiltonian of the system and $| \psi \rangle$ being a specific superposition of eigenstates of $H$ \cite{fvva}. The other is given in Ref. \cite{ml}, which states that the minimum time is given by $T_{\rm min} = \frac{\pi \hbar}{2(\bar E -E_0)}$, where $\bar E$ and $E_0$ are the mean energy and the lowest energy of the state which participates in the superposition.  Obviously, the results of Ref. \cite{fvva} and \cite{ml} will be equivalent if the condition $\Del E = \bar E -E_0$ is satisfied.  This condition 
can be simply satisfied by superposing two steady states homogeneously. 
According to the results in Refs. \cite{fvva, ml}, it is natural to assume that the minimum time should be
given by $T_{\rm min} = {\rm Max} \{\frac{\pi \hbar}{2 \Del E}, \ \frac{\pi \hbar}{2(\bar E -E_0)}\}$ \cite{glm}. 
A unified bound which contains both $\Delta E$ and $\bar E$ is considered \cite{lt}. 

An interesting connection between the minimum time of the quantum state evolving in Hilbert space and the average speed of an electron wave packet travelling  in spital space is constructed in a recent paper \cite{vd}. In this paper, the authors study  a relativistic electron coupling to a homogeneous magnetic field. 
They find that the average speed of this electron wave packet moving in  radial direction during the interval in which a quantum state evolving from an initial state to the orthogonal final one in Hilbert space is less than  the speed of light in vacuum, regardless of  intensities of the magnetic field one applies. It seems that, as expected,  Lorentz invariance is not violated  in this relativistic quantum mechanical model. Lorentz invariance would be violated in the non-relativistic limit of this model since the average speed of the electron wave packet in  radial direction exceeds  the speed of light in vacuum provided the intensity of the magnetic field is  strong enough. 

The work in Ref. \cite{vd} is generalized to noncommutative case recently. In the reference \cite{wj}, the authors study the noncommutative (both the coordinates and momenta are noncommutative simultaneously) Dirac equation. They find that Lorentz invariance will be violated in noncommutative Dirac equation since the average speed of an electron wave packet exceeds the speed of light in vacuum if the  magnetic field is strong enough.  In fact, the problem of violating Lorentz has been considered in Refs. \cite{jackiw, cai, Deriglazov1, Deriglazov2, Deriglazov3}. In Ref. \cite{jackiw}, the authors find that because of noncommutativity, Lorentz invariance will be violated in the noncommutative quantum electro-dynamics (QED) since the electromagnetic wave travels in different speed along different directions at the presence of a background magnetic field. A similar result is also obtained in Ref.\cite{cai}. This problem is also considered semiclassically from both noncommutative and gravitational
points of view \cite{Deriglazov1, Deriglazov2, Deriglazov3}. 

In this paper, we investigate the problems of whether Lorentz invariance is violated for  spin-0 or spin-1 relativistic bosons in two-dimensional spatial space. We  study the commutative case firstly and then generalize to the noncommutative case. The organization of this paper is as follows: In next section, we  study  spin-0 and spin-1 charged bosons coupling to  homogeneous magnetic fields on commutative plane. Then, in  section III, we  study  the noncommutative case. Some remarks and further discussions will be presented in the last section.

\section{Spin-0 and spin-1 charged bosons coupling to a homogenous magnetic field on commutative plane}

In this section, we  study spin-0 and spin-1 bosons coupling to a homogeneous magnetic field on a commutative plane. We start our studies from the spin-0 bosons.  


As stated before, the five-dimensional representation of the algebra (\ref{cbeta}) describes spin-0 boson. The explicit expressions of  five-dimensional $\beta^\mu$ matrices are
\begin{eqnarray}
\beta^0 = \left ( \begin{array}{cc} \Theta & \bar 0 \\ {\bar 0}^T& 0 \end{array} \right ), \ \beta^i = \left (\begin{array}{cc} \tilde 0 & \rho^i \\ (- \rho^i) ^T&0 \end{array} \right), \label{beta}
\end{eqnarray}
in which
\begin{eqnarray}
\Theta = \left (\begin{array}{cc}0&1\\1&0 \end{array} \right ), \ \rho^1 = \left( \begin{array}{ccc}-1&0&0\\0&0&0 \end{array} \right), \no \\
\rho^2 = \left( \begin{array}{ccc}0&-1&0\\0&0&0 \end{array} \right), \ \rho^3 = \left( \begin{array}{ccc}0&0&-1\\0&0&0 \end{array} \right).
\end{eqnarray}
Here $\bar 0$, $\tilde 0$ and $0$ are $2 \times 3$, $2 \times 2$ and $3 \times 3$ zero matrices respectively. Choosing symmetric gauge
$
A_i = - \half B \eps_{ij} x_j
$
and introduce  Larmor frequency
$
\omega_L = \frac{qB}{2m_0},
$
we write equation (\ref{dkp}) in the form
\begin{eqnarray}
\Big[ \beta^0 E & +&\beta^1 (p_1- m_0 \omega_L x_2 )  \\
&+& \beta^2 (p_2 + m_0 \omega_L x_1 ) -m_0  \Big ] \psi(x_1, x_2,t)=0, \label{dequation}
\end{eqnarray}
where $\psi(x_1,x_2,t)$ is a five-component wave function  
$\psi(x_1, x_2,t) = (\psi^1, \ \psi^2, \ \psi^3, \ \psi^4, \ \psi^5)^T (x_1, x_2,t)$.

Substituting the explicit expressions of $\beta^\mu$ matrices (\ref{beta}) into the above equation, we get a set of equations
\begin{subequations} \label{dkpeq}
\begin{align}
-m_0  \psi^1+ E \psi^2 - (p_1- m_0 \omega_L x_2 ) \psi^3 \no \\ - (p_2 + m_0 \omega_L x_1 ) \psi^4=0, \tag{\theequation a} \\
 E \psi^1 -m_0  \psi^2=0, \tag{\theequation b}\no \\
(p_1- m_0 \omega_L x_2) \psi_1 -m_0  \psi^3 =0, \tag{\theequation c} \\
(p_2+ m_0 \omega_L x_1) \psi_1 -m_0  \psi^4=0, \tag{\theequation d} \\
\psi^5 =0 \tag{\theequation e}.
\end{align}
\end{subequations}
Obviously, the five components $(\psi^1, \ \psi^2, \ \psi^3, \ \psi^4, \ \psi^5)$ are not independent each other.

Combining the above equations, we get the dynamical equation of the component $\psi^1$. It is
\begin{eqnarray}
(E^2 - m_0^2) \psi^1 = 2m_0 H_L \psi^1,  \label{dynamic}
\end{eqnarray}
where $H_L$ is the Landau Hamiltonian,
\begin{equation}
H_L = \frac{1}{2m_0} (p_1^2 + p^2_2) + \half m_0 \omega_L^2 (x_1 ^2 + x_2 ^2) + \omega_L L_z, \label{ha1}
\end{equation}
with $L_z = xp_y -y p_x$ being the angular momentum along $z$ direction.
Obviously, the Hamiltonian (\ref{ha1}) describes a planar non-relativistic charged particle interacting with a homogeneous perpendicular magnetic field.



The equation (\ref{dynamic}) is easily solved. The eigenvalues and eigenfunctions respectively are \cite{Yoshioka}
\begin{equation}
E_{n, m_l}= \pm \sqrt{m_0 ^2 + 2 m_0 (n+ m_l +1)\omega_L} \label{eigenvalues}
\end{equation}
and
\begin{eqnarray}
\psi^1 &=& F_{n,m_l} (r, \varphi) = \frac{{(-1)}^{\frac{n- |m_l|}{2}} (\frac{n- |m_l|}{2})!}{\sqrt {\pi (\frac{n+|m_l|}{2})! (\frac{n- |m_l|}{2})!} } \label{efs} \\
&& \times \alpha (\alpha r) ^ {|m_l|} L_{(\frac{n-|m_l|}{2})} ^{|m_l|} (\alpha^2 r^2) e ^{- \half \alpha ^2 r^2} e ^{i m_l \varphi}, \no
\end{eqnarray}
where $n=0,1,2,\cdots,  \ m_l = -n, -n+2, \cdots, n-2, n$, $\alpha = \sqrt {m_0 \omega_L}= \sqrt{\frac{qB}{2}}$ and $ L_{(\frac{n-|m_l|}{2})} ^{|m_l|} $ is Laguerre's polynomials.  Thus, the multi-component wave function $\psi$ can be obtained if one chooses a specific solution of the component $\psi^1$.

In order to simplify our further calculation and make a comparison with the spin-$\half$ case \cite{vd}, 
we choose two steady states whose first components respectively are (we only consider the positive-energy sector)
\begin{eqnarray}
\psi^1 = F_{0,0} (r, \varphi) = \frac{1}{\sqrt \pi} \alpha e ^{- \half \alpha^2 r^2}
\end{eqnarray}
and
\begin{eqnarray}
\psi^1 = F_{2,0} = - \frac{1}{\sqrt \pi} \alpha (1- \alpha^2 r^2) e^{- \half \alpha^2 r^2}.
\end{eqnarray}

Then,  after some direct calculations, we get these two steady states. They are
\begin{equation} \label{st00}
\phi_{0,0} = N_{0,0}  \left ( \begin{array}{c} F_{0,0} \\ \frac{E_{0,0}}{m_0} F_{0,0} \\  \frac{i \alpha}{m_0} F_{1,1} \\  \frac{\alpha}{m_0} F_{1,1} \\ 0 \end{array} \right ) e^{- i{E_{0,0}}t},
\end{equation}
and
\begin{equation} \label{st20}
\phi_{2,0} = N_{2,0}  \left ( \begin{array}{c} F_{2,0} \\ \frac{E_{2,0}}{m_0} F_{2,0} \\  \frac{i \alpha}{m_0} (\sqrt 2 F_{3,1} - F_{1,-1}) \\  \frac{\alpha}{m_0} (F_{3,1} +  F_{1,-1}) \\ 0 \end{array} \right )  e^{- i {E_{2,0}}t},
\end{equation}
where
\begin{equation}
N_{0,0} = \frac{m_0}{\sqrt{ m_0 ^2 + E_{0,0}^2 + 2 \alpha^2}}, \ N_{2,0} = \frac{m_0}{\sqrt {m_0 ^2 + E_{2,0}^2 + 6 \alpha^2}} \label{nconstant}
\end{equation}
are two normalization constants.

For the purpose of avoiding the controversy on the minimum time for a quantum state evolving in the Hilbert space from an initial state to the orthogonal final one is determined by the energy variance $\Del E$ or the mean energy $\bar E$, we  superpose these two steady states (\ref{st00}, \ref{st20}) homogeneously, i.e.,
\begin{equation}
\Psi(r, \varphi, t) = \frac{1}{\sqrt 2} \big[ \phi_{0,0}(r, \varphi) e ^{ - {i}{ E_{0,0}}t} + \phi_{2,0} (r, \varphi) e ^{ - {i} {E_{2,0}}t} \big]. \label{sss}
\end{equation}

According to Refs. \cite{fvva, ml}, the minimum time for a the state $\Psi(r, \varphi, t)$ evolving from the initial state $\Psi(r, \varphi, 0)$ to the final orthogonal state $\Psi(r, \varphi, T_{\rm min})$ is given by
$T_{\rm min} = \frac{\pi \hbar}{2 (\bar E - E_{0,0})}$, or, equivalnetly, $ T_{\rm min} = \frac{\pi \hbar}{2 \Del E}$, 
where $\bar E$ and 
$ \Del E$ are the mean energy and energy variance on the state $|\Psi (t) \rangle$ respectively. After direct calculation, we get
\begin{equation}
T_{\rm min}= \frac{\pi}{\sqrt{m_0 ^2 + 3 q B} - \sqrt{m_0^2 + qB}}. \label{cmt}
\end{equation}

The average radial displacement of the spinless boson in the interval $[0, \ T_{\rm min}]$ is given by
\begin{eqnarray}
\Del r &=& \Big | \langle \Psi(T_{\rm min})| r|\Psi(T_{\rm min}) \rangle - \langle \Psi(0)| r|\Psi0) \rangle \Big | \no \\
&=& 2 \Big | \langle \phi_{0,0}|r| \phi_{2,0} \rangle \Big |. \label{displacement}
\end{eqnarray}
Substituting (\ref{st00}, \ref{st20}, \ref{sss}) into the above equation and after some direct calculations, we arrive at
\begin{equation}
\Del r = (\frac{1}{4} + \frac{3 \sqrt 3}{8}) \frac{\sqrt \pi}{\alpha}.
\end{equation}
Thus, the average speed of the wave-packet of this charged boson along the radial direction during this interval is
\begin{eqnarray}
\bar v =\frac{\Del r}{T_{\rm min}} &=& \frac{2 + 3 \sqrt 3}{8} \sqrt{\frac{2}{\pi q B} }(\sqrt{m_0^2 + 3 qB }- \sqrt{m_0^2 + qB}).
\label{aspeed}
\end{eqnarray}
The average speed $\bar v$ is a monotone increasing function of $B$. The  average speed $\bar v$ reaches its maximum when $B \to \infty$. It is
\begin{equation}
{\bar v}_{\rm max}= \lim_{B \to \infty} \bar v = \frac{2 + 3 \sqrt 3}{8}\frac{(\sqrt 6 -\sqrt 2)}{\sqrt \pi} \doteq 0.5254 <1. \label{maxs}
\end{equation}
It shows that the average radial speed of this spinless boson wave-packet is less than  the speed of light in vacuum ($c=1$) no matter the intensities of  the magnetic field. Therefore, Lorentz invariance is not violated in this relativistic quantum mechanics model.

Now, we study the DKP equation with $10$-dimensional representation of the algebra (\ref{cbeta}). As is  known, it describes spin-1 bosons. The explicit expressions of $\beta$ matrices are
\begin{equation}
\beta^0 = \left( \begin{array}{cccc}0& \check 0& \check 0& \check 0 \\{\check 0} ^T& 0_{3 \times 3} &I_{3 \times 3} &0_{3 \times 3} \\ {\check 0} ^T& I_{3 \times 3}& 0_{3 \times 3}&0_{3 \times 3} \\ {\check 0} ^T& 0_{3 \times 3} &0_{3 \times 3} &0_{3 \times 3} \end{array} \right)
\end{equation}
and
\begin{equation}
\beta^i = \left( \begin{array}{cccc}0& \check 0& e_i& \check 0 \\ {\check 0} ^T& 0_{3 \times 3} &0_{3 \times 3} &-i S_i \\- {e_i}^T &0_{3 \times 3}&0_{3 \times 3}&0_{3 \times 3} \\{\check 0} ^T& - iS_i &0_{3 \times 3} &0_{3 \times 3} \end{array} \right),
\end{equation}
in which
\begin{equation}
\check 0 = (0,0,0), \ e_1 =(1,0,0), \ e_2 =(0,1,0), \ e_3 =(0,0,1),
\end{equation}
and
\begin{eqnarray}
S_1 = \left(\begin{array}{ccc}0&0&0 \\0&0&-i \\0&i&0 \end{array} \right), \ S_2 =  \left(\begin{array}{ccc}0&0&i \\0&0&0 \\-i&0&0 \end{array} \right), S_3 = \left(\begin{array}{ccc}0&-i&0 \\i&0&0 \\0&0&0 \end{array} \right).
\end{eqnarray}

Using the explicit expressions of $\beta$ matrices, we write the DKP equation in the component form,
\begin{subequations} \label{sp1dkp}
\begin{align}
(p_1 - m_0\omega_L x_2) \psi^5 + (p_2 + m_0 \omega_L x_1) \psi^6 = m_0\psi^1,  \label{spldkp1} \\
E \psi^5 + (p_2 + m_0\omega_L x_1) \psi^{10} = m_0 \psi^2, \label{spldkp2} \\
E \psi^6 - (p_1 - m_0\omega_L x_2) \psi^{10} = m_0 \psi^3,  \label{spldkp3}  \\
E \psi^7 - (p_2 + m_0\omega_L x_1) \psi^{8} + (p_1 - m_0\omega_L x_2) \psi^9  = m_0 \psi^4, \label{spldkp4}  \\
E \psi^2 - (p_1 - m_0\omega_L x_2) \psi^{1} = m_0 \psi^5, \label{spldkp5} \\
E \psi^3 - (p_2 + m_0\omega_L x_1) \psi^{1} = m_0 \psi^6, \label{spldkp6} \\
E \psi^4 = m_0 \psi^7, \label{spldkp7} \\
 (p_2 + m_0\omega_L x_1) \psi^{4} = m_0 \psi^8, \label{spldkp8} \\
-(p_1 - m_0\omega_L x_1) \psi^{4} = m_0 \psi^9, \label{spldkp9} \\
-(p_2 + m_0\omega_L x_1) \psi^{2} + (p_1 - m_0\omega_L x_2) \psi^3  = m_0 \psi^{10}. \label{spldkp10}
\end{align}
\end{subequations}
According to equations (\ref{spldkp4}, \ref{spldkp7}, \ref{spldkp8}, \ref{spldkp9}), we get the dynamical equation of the component $\psi^4$ 
\begin{eqnarray}
(E^2 - m_0^2) \psi^4 = 2m_0 H_L \psi^4,  \label{dynamic1}
\end{eqnarray}
where $H_L$ is given in (\ref{ha1}).

Therefore, the solutions of $\psi^4$ in spin-1 case take the same form as (\ref{eigenvalues}, {\ref{efs}), i.e.,
\begin{equation}
E_{n, m_l} = \pm \sqrt{ m_0 ^2 + 2 m_0 (n + m_l +1) \omega_L}
\end{equation}
and
\begin{eqnarray}
\psi^4 &=& F_{n,m_l} (r, \varphi) = \frac{{(-1)}^{\frac{n- |m_l|}{2}} (\frac{n- |m_l|}{2})!}{\sqrt {\pi (\frac{n+|m_l|}{2})! (\frac{n- |m_l|}{2})!} } \label{efs1} \\
&& \times \alpha  (\alpha   r) ^ {|m_l|} L_{(\frac{n-|m_l|}{2})} ^{|m_l|} (\alpha^{  2} r^2) e ^{- \half \alpha ^{  2} r^2} e ^{i m_l \varphi}. \no
\end{eqnarray}
Following Ref. \cite{iran}, 
we choose a special solution for  components $\psi^2$ and $\psi^3$ 
\begin{equation}
\psi^2 =0 \quad {\rm and } \quad \psi^3=0. \label{sc}
\end{equation}
As a result, the components $\psi^1=\psi^5= \psi^6=0$. Thus, the ten-component eigenfunction can be obtained if we select a specific solution for $\psi^4$ from (\ref{efs1}). In order to simplify our future calculations and make a comparison with Ref. \cite{vd}, we choose two specific solutions for $\psi^4$. They are
\begin{equation}
\psi^4 = F_{0,0}=  \frac{1}{\sqrt \pi} \alpha e ^{- \half \alpha^2 r^2}
\end{equation}
and
\begin{equation}
\psi^4 = F_{2,0} =  - \frac{1}{\sqrt \pi} \alpha  (1- \alpha^{  2} r^2) e^{- \half \alpha^{  2} r^2}.
\end{equation}
Thus, the explicit expressions for two steady states we want to superpose are
\begin{equation}
\phi_{0,0} = N_{0,0}\left( \begin{array}{c}0 \\0 \\ 0\\ F_{0,0} \\0\\0\\ \frac{E_{0,0}}{m_0} F_{0,0} \\  \frac{\alpha }{m_0} F_{1,1} \\ -\frac{i \alpha }{m_0} F_{1,1} \\0 \end{array} \right), \label{spin1st1}
\end{equation}
and
\begin{equation}
\phi_{2,0} = N_{2,0} \left( \begin{array}{c}0\\0\\0\\F_{2,0}\\0\\0\\ \frac{E_{2,0}}{m_0} F_{2,0} \\\frac{\alpha }{m_0} (\sqrt 2 F_{3,1} + F_{1,-1}) \\ -\frac{i \alpha }{m_0} (\sqrt 2 F_{3,1} - F_{1,-1}) \\0 \end{array} \right), \label{spin1st2}
\end{equation}
where  $N_{0,0}, \ N_{2,0}$ are identical to (\ref{nconstant}).  

We superpose these two steady states (\ref{spin1st1}, \ref{spin1st2}) homogeneously
\begin{equation}
\Psi (r, \varphi, t) = \frac{1}{\sqrt 2} [\phi_{0,0}(r, \varphi) e^{-i{E_{0,0}} t} + \phi_{2,0}(r, \varphi) e^{-i{E_{2,0}}t} ]. \label{spin1sus}
\end{equation}
The eigenvalues are equal to the ones in spin-0 case, thus the minimum time for state (\ref{spin1sus}) evolving from the initial state $\Psi(r, \varphi, 0)$ to the final orthogonal state $\Psi(r, \varphi, T_{\rm min})$ is the same as (\ref{cmt}). The radial displacement of the wave packet during the interval  $[0, \ T_{\rm min}]$ can be calculated straightforwardly. The result is nothing but (\ref{displacement}). Therefore, the average velocity along radial direction of this spin-1 boson is the same as in (\ref{aspeed}). The maximum average speed is achieved when the intensity of the magnetic field tends to infinity. The result is identical to (\ref{maxs}). It is less than  the speed of light in vacuum. Thus, it shows that Lorentz invariance is not violated.


\section{Spin-0 and spin-1 bosons on noncommutative plane}

We  generalize our previous studies to the noncommutative plane in this section.  Noncommutativity has attracted much attention due to superstring theories in recent years \cite{string1, string2, string3, string4}. There are a large number of  papers studying quantum field theories in noncommutative space \cite{nqf1,nqf2,nqf3,nqf4,nqf5,nqf6}. 
Non-relativistic noncommutative quantum mechanical models, such as noncommutative harmonic oscillator, noncommutative Landau problem have been studied extensively \cite{ncqm1,ncqm2,ncqm3,ncqm4,ncqm5,ncqm6,ncqm7}. 
The relativistic quantum mechanical models on noncommutative space are also investigated since the work of \cite{ndha}. Some geometrical phases in noncommutative relativistic quantum theory are studied  Refs. \cite{ma1, ma2, ma3} recently. Interestingly, it is found  that the noncommutativity even has some relationship with JC model in quantum optics context \cite{optics}. 


As we have shown, the average speeds of the wave packets of  spin-0 and spin-1  bosons  described by DKP equation along the radial direction will not exceed the speed of light in vacuum, regardless of the intensity of the magnetic field.  
A natural question is: does the Lorentz invariance be violated in the noncommutative DKP equation? In the following, we will investigate DKP equation in noncommutative $2+1$-dimensional phase space.

The noncommutative $2+1$-dimensional phase space is described by the algebraic relation
\begin{eqnarray}
\lbrack \hat x_i, \ \hat x_j ] &=& i \theta \eps_{ij}, \quad [\hat p_i, \ \hat p_j ] = i \eta \eps_{ij}, \no \\
\lbrack \hat x_i, \ \hat p_j ] &=& i (1 + \frac{\theta \eta}{4}) \del_{ij} , \ \ \ \ \ \ \quad i, j =1,2, \label{ncps2}
\end{eqnarray}
where 
$\theta$ and $ \eta$ are two real parameters, and $\eps_{ij}$ is the 2-dimensional anti-symmetric tensor.
In order to avoid the problem of  unitarity, we only consider  noncommutativities among coordinates and momenta. 

The most general method  of studying  noncommutative quantum mechanics is to  assume that the dynamical equations in noncommutative quantum mechanics take the same form as their commutative counterparts. However,  variables in dynamical equation are replaced by the corresponding noncommutative ones. Therefore, the noncommutative version of spin-0 DKP equations (\ref{dkpeq}) is
\begin{subequations}
\begin{align}
-m_0  \psi^1+ E \psi^2 - (\hat p_1- m_0 \omega_L \hat x_2 ) \psi^3 \no \\ - (\hat p_2 + m_0 \omega_L \hat x_1 ) \psi^4=0, \label{ncdkpeqa}  \\
E \psi^1 -m_0\psi^2 =0, \label{ncdkpeqb} \\
-m_0  \psi^3 + (\hat p_1- m_0 \omega_L \hat x_2) \psi_1 =0, \label{ncdkpeqc}   \\
-m_0  \psi^4 + (\hat p_2+ m_0 \omega_L \hat x_1) \psi_1 =0, \label{ncdkpeqd} \\
\psi^5 =0. \label{ncdkpeqe}
\end{align}
\end{subequations}
In which $(\hat x_i, \ \hat p_i)$ satisfy the algebraic relation (\ref{ncps2}) and we have  chosen the symmetric gauge  $\hat A_i = - \frac{B}{2} \eps_{ij} \hat x_j$ \cite{zhangjz}.

Combining the above equations, we get
\begin{eqnarray}
(E^2 - m_0^2) \psi^1 = 2m_0 \hat H_L^{NC} \psi^1,  \label{ndynamic}
\end{eqnarray}
where 
\begin{equation}
\hat H_L ^{NC}= \frac{1}{2m_0} (\hat p_1^2 + \hat p^2_2) + \half m_0 \omega_L^2 (\hat x_1 ^2 + \hat x_2 ^2) + \omega_L \hat L_z, \label{nha}
\end{equation}
with $\hat L_z = \hat x \hat p_y - \hat y \hat p_x$ being the noncommutative angular momentum along $z$ direction.
Hamiltonian (\ref{nha}) 
describes a planar non-relativistic charged particle interacting with a homogeneous perpendicular magnetic field on the noncommutative plane (\ref{ncps2}).

We map  noncommutative variables $(\hat x_i, \ \hat p_i)$  to  commutative ones $( x_i, \  p_i)$ which satisfy the standard Heisenberg algebra
\begin{equation}
[x_i, \ x_j]=[p_i, \ p_j]=0, \quad [x_i, \ p_j] = i  \del_{ij}.
\end{equation}
It is straightforward to check that the map from noncommutative variables to commutative ones can be realized by \cite{bertolami}
\begin{equation}
\hat x_i = x_i - \frac{\theta}{2 } \eps_{ij} p_j, \quad \hat p_i = p_i + \frac{\eta}{2 } \eps_{ij} x_j. \label{cvs}
\end{equation}

In terms of  commutative variables $(x_i, \ p_i)$, we find the dynamical equation for  component $\psi^1$ from equations (\ref{ncdkpeqa}, \ref{ncdkpeqb}, \ref{ncdkpeqc}, \ref{ncdkpeqd}),
\begin{equation}
(E^2 - m_0^2) \psi^1 = 2m_{0} H_L^{NC} \psi^1, \label{ncdynamic1}
\end{equation}
where $H_L^{NC}$ is the noncommutative Landau Hamiltonian expressed in terms of commutative variables $(x_i, \ p_i)$. It is
\begin{eqnarray}
H_L ^{NC} = \frac{1}{2 M_{\rm eff} } (p_1 ^2 + p_2 ^2) + \half M_{\rm eff} \Omega ^2 (x_1 ^2 + x_2 ^2)+ \Omega L_z, \label{nonha1}
\end{eqnarray}
in which
\begin{equation}
M_{\rm eff} = \frac{m_0}{(1- \frac{q \theta B}{4})^2}, \quad \Omega^2 = \bigg [\frac{1}{m_0} (1- \frac{q \theta B}{4}) ( \frac{qB}{2}-\frac{\eta}{2}) \bigg]^2 .
\end{equation}


The eigenvalues and eigenfunctions of equation (\ref{ncdynamic1}) are
\begin{eqnarray}
E_{n, m_l} &=& \pm \sqrt{m_0 ^2 + 2 m_0 (n+ m_l +1) \Omega} \label{ncss0}
\end{eqnarray}
and
\begin{eqnarray}
\psi^1 &=& F_{n,m_l} (r, \varphi) = \frac{{(-1)}^{\frac{n- |m_l|}{2}} (\frac{n- |m_l|}{2})!}{\sqrt {\pi (\frac{n+|m_l|}{2})! (\frac{n- |m_l|}{2})!} } \label{nefs} \\
&& \times \alpha^\prime (\alpha^\prime r) ^ {|m_l|} L_{(\frac{n-|m_l|}{2})} ^{|m_l|} (\alpha^{\prime 2} r^2) e ^{- \half \alpha ^{\prime 2} r^2} e ^{i m_l \varphi}, \no
\end{eqnarray}
where $n=0,1,2,\cdots, \ m_l= -n, -n+2, \cdots, n-2, n$, and $$\alpha^\prime =\sqrt{M_{\rm eff} \Omega}$$
Other components can be calculated directly from the commutative version of equations (\ref{ncdkpeqb}, \ref{ncdkpeqc}, \ref{ncdkpeqd})
\begin{subequations}
\begin{align}
E \psi^1- m_0 \psi^2 =0, \label{ncdynamic2} \\
-m_0 \psi^3 + \big [(1- \frac{q \theta B}{4})p_1 +  (\frac{\eta}{2} - \frac{qB}{2}) x_2 \big ] \psi^1 =0 , \label{ncdynamic3} \\
-m_0 \psi^4 + \big [(1- \frac{q \theta B}{4})p_2 -  (\frac{\eta}{2} - \frac{qB}{2}) x_1 \big ] \psi^1 =0 . \label{ncdynamic4}
\end{align}
\end{subequations}

In order to make a comparison with  commutative version, we choose  two solutions of $\psi^1$ as
\begin{eqnarray}
\psi^1 = F_{0,0} = \frac{1}{\sqrt \pi}\alpha^\prime e^{- \half \alpha^{\prime2} r^2}
\end{eqnarray}
and
\begin{equation}
\psi^1 = F_{2,0} = -\frac{1}{\sqrt \pi} \alpha^\prime (1- \alpha^{\prime 2} r^2) e^{- \half \alpha^{\prime2} r^2}.
\end{equation}
Then, the corresponding two steady states are
\begin{equation}
\phi_{0,0} =  N_{0,0}  \left ( \begin{array}{c} F_{0,0} \\ \frac{E_{0,0}}{m_0} F_{0,0} \\  i \frac{ \alpha ^\prime}{m_0} (1- \frac{q \theta B}{4})  F_{1,1} \\  \frac{\alpha \prime}{m_0}(1- \frac{q \theta B}{4}) F_{1,1} \\ 0 \end{array} \right ) \label{ncst00}
\end{equation}
and
\begin{equation}
\phi_{2,0} = N_{2,0}  \left ( \begin{array}{c} F_{2,0} \\ \frac{E_{2,0}}{m_0} F_{2,0} \\ i \frac{\alpha^\prime}{m_0} (1- \frac{q \theta B}{4})  (\sqrt 2 F_{3,1} - F_{1,-1}) \\  \frac{\alpha^\prime}{m_0} (1- \frac{q \theta B}{4})  (\sqrt 2F_{3,1} +  F_{1,-1}) \\ 0 \end{array} \right ), \label{ncst20}
\end{equation}
where $N_{0,0}$ and $N_{2,0}$ are two normalization constants. They are
\begin{equation}
N_{0,0} = \frac{m_0}{\sqrt {{m_0 ^2 + E_{0,0}^2 + 2 \alpha^{\prime 2} (1- \frac{q \theta B}{4})^2}}}
\end{equation}
and
\begin{equation}
N_{2,0} = \frac{m_0} {\sqrt {{m_0 ^2 + E_{2,0}^2 + 6 \alpha^{\prime 2} (1- \frac{q \theta B}{4})^2}}}
\end{equation}
respectively.

We superpose two steady states (\ref{ncst00}) and (\ref{ncst20}) homogeneously. Thus the state we prepared is
\begin{equation}
\Psi (r, \varphi,t)= \frac{1}{\sqrt 2} \big [\phi_{0,0}(r, \varphi) e^{-i {E_{0,0}}t} + \phi_{2,0}(r, \varphi)  e^{-i { E_{2,0}} t} \big ].  \label{ncstate}
\end{equation}
According to Refs. \cite{fvva,ml}, the minimum time for  state (\ref{ncstate}) evolving from the initial state $\Psi(r, \varphi, 0)$ to the final orthogonal one $\Phi(r, \varphi, T_{\rm min})$ is given by $T_{\rm min} = \frac{\pi \hbar}{2(\bar E -E_{0,0})}= \frac{\pi \hbar}{2 \Del E}$. Substituting the eigenvalues (\ref{ncss0}) for states $\phi_{0,0}$ and $\phi_{2,0}$ into the expression of the minimum time, we get
\begin{eqnarray}
T_{\rm min}= \frac{\pi}{\sqrt{m_0^2 + 6 m_0 \Omega}- \sqrt{m_0^2 + 2 m_0 \Omega}}.\label{ncmt}
\end{eqnarray}
Accordingly, the displacement along the radial direction during the period of time $T_{\rm min}$ is given by the expression $\Del r =\big |\langle \Psi(T_{\rm min}) |r |\Psi(T_{\rm min}) \rangle  - \langle \Psi({0}) |r |\Psi({0}) \rangle \big | = 2\big |\langle \phi_{0,0} |r |\phi_{2,0} \rangle \big | $. Substituting equations (\ref{ncstate}) into the expressions of $\Del r$ and taking the limit of $B \to \infty$, we get
\begin{equation}
\Del r = \big |\langle \phi_{0,0} |r |\phi_{2,0} \rangle \big | =\frac{2 + 3 \sqrt 3}{8} \alpha^\prime. \label{nondisplacement}
\end{equation}
Therefore, the average speed along the radial direction of wave packet during the interval $[0, \ T_{\rm min}]$ in the limit $B \to \infty$ is given by
\begin{equation}
\bar v = \frac{\Delta r}{T_{\rm min}} \doteq 0.5254\big|1- \frac{q \theta B}{4} \big|. \label{nv}
\end{equation}

Compared with the commutative case, we find that due to spatial noncommutativity, there is an extra factor $\big | 1 - \frac{q\theta B}{4} \big |$. It is this factor which enables average radial speed of the wave packet to exceed  the speed of light in vacuum provided the intensity of  the magnetic field is strong enough. 
It is a clear evidence of violating Lorentz invariance  in this noncommutative relativistic quantum mechanical model.

Now,  we study the noncommutative spin-1 DKP equation. In noncommutative plane, the dynamical equations take the same form as Eqs. (\ref{sp1dkp}) except  variables $(x_i, \ p_i)$ are replaced by  noncommutative ones $(\hat x_i, \ \hat p_i)$. Thus, the explicit form of the dynamical equations in noncommutative plane are
\begin{subequations} \label{sp1ncdkp}
\begin{align}
 (\hat p_1 - m_0\omega_L \hat x_2) \psi^5  + (\hat p_2 + m_0 \omega_L \hat x_1) \psi^6 = m_0\psi^1,  \label{splncdkp1} \\
E \psi^5 + (\hat p_2 + m_0\omega_L \hat x_1) \psi^{10} = m_0 \psi^2, \label{splncdkp2} \\
E \psi^6 - (\hat p_1 - m_0\omega_L \hat x_2) \psi^{10} = m_0 \psi^3,  \label{splncdkp3}  \\
E \psi^7 - (\hat p_2 + m_0\omega_L \hat x_1) \psi^{8} + (\hat p_1 - m_0\omega_L \hat x_2) \psi^9  = m_0 \psi^4, \label{splncdkp4}  \\
E \psi^2 - (\hat p_1 - m_0\omega_L \hat x_2) \psi^{1} = m_0 \psi^5, \label{splncdkp5} \\
E \psi^3 - (\hat p_2 + m_0\omega_L \hat x_1) \psi^{1} = m_0 \psi^6, \label{splncdkp6} \\
E \psi^4 = m_0 \psi^7, \label{splncdkp7} \\
(\hat p_2 + m_0\omega_L \hat x_1) \psi^{4} = m_0 \psi^8, \label{splncdkp8} \\
-(\hat p_1 - m_0\omega_L \hat x_2) \psi^{4} = m_0 \psi^9, \label{splncdkp9} \\
-(\hat p_2 + m_0\omega_L \hat x_1) \psi^{2} + (\hat p_1 - m_0\omega_L \hat x_2) \psi^3  = m _0\psi^{10}. \label{splncdkp10}
\end{align}
\end{subequations}
Mapping  noncommutative variables $(\hat x_i, \ \hat p_i)$ to  commutative ones $(x_i, \ p_i)$, we get the commutative version of Eqs. (\ref{sp1ncdkp}),
\begin{subequations} \label{sp1ncdkpp}
\begin{align}
 \big[(1 - \frac{q \theta B}{4})  p_1 + (\frac{\eta}{2} - \frac{q B}{2}) x_2 \big] \psi^5 \no  \\+ \big[(1 - \frac{q \theta B}{4})  p_2 - (\frac{\eta}{2} - \frac{q B}{2}) x_1 \big ]\psi^6 = m_0\psi^1,  \label{splncdkp1p} \\
E \psi^5 + \big[(1 - \frac{q \theta B}{4})  p_2 - (\frac{\eta}{2} - \frac{q B}{2}) x_1 \big ] \psi^{10} = m_0 \psi^2, \label{splncdkp2p} \\
E \psi^6 - \big[(1 - \frac{q \theta B}{4})  p_1 + (\frac{\eta}{2} - \frac{q B}{2}) x_2 \big]\psi^{10} = m _0\psi^3,  \label{splncdkp3p}  \\
E \psi^7 - \big[(1 - \frac{q \theta B}{4})  p_2 - (\frac{\eta}{2} - \frac{q B}{2}) x_1 \big ] \psi^{8}\no\\
+\big[(1 - \frac{q \theta B}{4})  p_1 + (\frac{\eta}{2} - \frac{q B}{2}) x_2 \big] \psi^9  = m_0 \psi^4, \label{splncdkp4p}  \\
E \psi^2 - \big[(1 - \frac{q \theta B}{4})  p_1 + (\frac{\eta}{2} - \frac{q B}{2}) x_2 \big]  \psi^{1} = m_0 \psi^5, \label{splncdkp5p} \\
E \psi^3 - \big[(1 - \frac{q \theta B}{4})  p_2 - (\frac{\eta}{2} - \frac{q B}{2}) x_1 \big ] \psi^{1} = m_0 \psi^6, \label{splncdkp6p} \\
E \psi^4 = m_0 \psi^7, \label{splncdkp7p} \\
\big[(1 - \frac{q \theta B}{4})  p_2 - (\frac{\eta}{2} - \frac{q B}{2}) x_1 \big ] \psi^{4} = m_0 \psi^8, \label{splncdkp8p} \\
-\big[(1 - \frac{q \theta B}{4})  p_1 + (\frac{\eta}{2} - \frac{q B}{2}) x_2 \big] \psi^{4} = m_0 \psi^9, \label{splncdkp9p} \\
-\big[(1 - \frac{q \theta B}{4})  p_2 - (\frac{\eta}{2} - \frac{q B}{2}) x_1 \big ] \psi^{2}\no\\
 +  \big[(1 - \frac{q \theta B}{4})  p_1 + (\frac{\eta}{2} - \frac{q B}{2}) x_2 \big]  \psi^3  = m_0 \psi^{10}. \label{splncdkp10p}
\end{align}
\end{subequations}

The dynamical equation for component $\psi^4$ can be obtained from above equations. It is
\begin{equation}
(E^2 - m_0^2) \psi^4 = 2m_{0} H_L^{NC} \psi^4, \label{ncdynamic23}
\end{equation}
where $H_L^{NC}$ is given in (\ref{nonha1}). With the help of solutions to the equation (\ref{ncdynamic1}), 
we  get the solutions to the equation (\ref{ncdynamic23}) easier. The eigenvalues are identical to (\ref{ncss0}) and the corresponding solutions of component $\psi^4$ are identical with the solutions of $\psi^1$ in noncommutative spin-0 case (\ref{nefs}).

Similar with the commutative version, we set $\psi^2 =\psi^3 =0$. Then the other components can be determined by the equations (\ref{sp1ncdkpp}). For the sake of comparing with the corresponding commutative case, we choose two specific solutions of $\psi^4$,
\begin{eqnarray}
\psi^4 &=&  F_{0,0} = \frac{1}{\sqrt \pi} \alpha^\prime e^{- \half \alpha^{\prime2} r^2}, \no \\
\psi^4 &=& F_{2,0} = -\frac{1}{\sqrt \pi} \alpha^\prime (1- \alpha^{\prime 2} r^2) e^{- \half \alpha^{\prime2} r^2}.
\end{eqnarray}

Then, the two steady states we prepared are
\begin{equation}
\phi_{0,0} = N_{0,0} \left( \begin{array}{c}0\\0\\0\\F_{0,0}\\0\\0\\ \frac{E_{0,0}}{m_0} F_{0,0} \\  \frac{\alpha ^\prime}{m_0}(1 - \frac{q \theta B}{4} ) F_{1,1}  \\ -\frac{i\alpha ^\prime}{m_0}(1 - \frac{q \theta B}{4} ) F_{1,1} \\0 \end{array} \right) \label{ncspin1st1}
\end{equation}
and
\begin{equation}
\phi_{2,0} =N_{2,0} \left (\begin{array}{c}0\\ 0\\ 0\\ F_{2,0} \\0\\ 0\\ \frac{E_{2,0}}{m_0} F_{2,0} \\  \frac{\alpha}{m_0}(1 - \frac{q \theta B}{4} )(\sqrt 2 F_{3,1} + F_{1,-1}) \\-\frac{i \alpha}{m_0}(1 - \frac{q \theta B}{4} )(\sqrt 2 F_{3,1} - F_{1,-1}) \\0 \end{array} \right). \label{ncspin1st2}
\end{equation}

Superposing these two steady states homogeneously, we get the state
\begin{equation}
\Psi(r, \varphi, t) = \frac{1}{\sqrt 2} [ \phi_{0,0} (r, \varphi) e^{-i {E_{0,0}} t} + \phi_{2,0}(r, \varphi)  e^{-i {E_{2,0}}t}].
\end{equation}
The minimum time for this state evolving from the initial state $\Psi(r, \varphi, 0)$ to the orthogonal final one $\Psi(r, \varphi, T_{\rm min})$ is identical to (\ref{ncmt}). The displacement along the radial direction can also be calculated directly. It is equivalent to (\ref{nondisplacement}) in the limit of $B \to \infty$. Thus, the average speed of the wave packet during interval $[0, \ T_{\rm min}]$ is given by (\ref{nv}). Compared with the commutative counterpart, there is an extra factor $|1- \frac{q \theta B}{4}|$ which enables the average speed along the radial direction to exceed light in vacuum.  Therefore, it means that  Lorentz invariance is also violated in this relativistic spin-1 boson due to noncommutativity. 

\section{conclusions and remarks}

In this paper, we investigate the problem of whether Lorentz invariance is violated in noncommutative quantum mechanics regime. In fact, the violation of Lorentz invariance due to noncommutativity has been noticed more than 10 years in Refs. \cite{jackiw, cai}. Nevertheless, there is a  difference between Refs. \cite{jackiw, cai}  and ours. In Refs. \cite{jackiw, cai}, the authors investigate the propagation of electromagnetic wave in noncommutative space. The electromagnetic wave, from the quantum point of view, is photons, which are massless.

We study the charged  massive spin-0 and spin-1 relativistic bosons in the presence of homogeneous magnetic fields in $2$-dimensional space. Both commutative and noncommutative cases are studied. According to the theory of special relativity, the speed of a massive particle can not exceed the speed of light in vacuum. In our studies, we find 
that the average radial speeds of wave packets during the interval $[0, \ T_{\rm min}]$ in commutative plane are less than the speed of light in vacuum, no matter how strong the intensity of the magnetic field is.  However, when noncommutativity is taken into account, we find that the average radial speed will exceed  the speed of light in vacuum if the intensity of the magnetic field is strong enough. 
It conflicts with  the  special relativity directly. Therefore, it indicates that Lorentz invariance will be violated in noncommutative space in quantum mechanics regime.

It is known that there are redundant degrees of freedom in DKP equation. For the spin-0 case, the physical degree of freedom is easy to obtained. However, it is not  straightforward to get the degrees of freedom for spin-1 case \cite{cc}. The solutions (\ref{spin1st1}, \ref{spin1st2}) and (\ref{ncspin1st1}, \ref{ncspin1st2}) are based on $\psi^2 = \psi^3 =0$, which  are special solutions to DKP equation. It may be worthwhile to study  whether the same conclusions still be held for the general solutions of the DKP equation  \cite{wj2}.

\section*{Acknowledgement}
This work is supported by the NSFC with Grant No 11465006.

\end{document}